\def\draftversion{false}
\RequirePackage{ifthen}
\ifthenelse{\equal{\draftversion}{false}}{
  \documentclass[citeautoscript,floatfix,aps,prl,twocolumn,preprintnumbers,
      superscriptaddress,longbibliography,amsmath,amssymb]{revtex4-2}
}{
  \documentclass[citeautoscript,floatfix,aps,prx,twocolumn,
      superscriptaddress,longbibliography]{revtex4-2}
}

\usepackage{amsmath}
\usepackage{graphicx}
\usepackage{epstopdf}
\usepackage{natbib}
\usepackage{array}
\usepackage{dcolumn}% Align table columns on decimal point
\usepackage{bm}% bold math

\usepackage{soul}  % \st    for strike-out

\usepackage[usenames,dvipsnames]{color}

\newcommand\T{\rule{0pt}{2.6ex}}              % Top strut
\newcommand\B{\rule[-1.2ex]{0pt}{0pt}}        % Bottom strut

\soulregister\cite7

\ifthenelse{\equal{\draftversion}{true}}{
  \marginparwidth 2.7in
  \marginparsep 0.5in
  \newcounter{comm} % counter for commentaries
  \def\commnext{\stepcounter{comm}}
  \def\commtext{{\bf\color{blue}[\arabic{comm}]}}
  \def\commmar{{\bf\color{blue}[\arabic{comm}]}}
  \def\dvm#1{\commnext\marginpar{\small DV\commmar: #1}\commtext}
  \def\cdm#1{\commnext\marginpar{\small CED\commmar: #1}\commtext}
  \def\msm#1{\commnext\marginpar{\small MS\commmar: #1}\commtext}
  \def\asm#1{\commnext\marginpar{\small AS\commmar: #1}\commtext}
  \def\miq#1{\commnext\marginpar{\small MR\commmar: #1}\commtext}
  \def\mlab#1{\marginpar{\small\bf #1}}
  
}{
  \def\dvm#1{}
  \def\cdm#1{}
  \def\msm#1{}
  \def\asm#1{}
  \def\miq#1{}
  \def\mlab#1{}
  
}
\def\E{\mathcal{E}}

\def\Ea{{\mathcal{E}_a}}

\newcommand{\w}{\omega}

\newcommand{\e}{\epsilon}

\begin{document}

\title{ {Rototranslational sum rules for {nuclear dynamics} via traveling pseudopotentials}}
%Pseudopotentials that move: From inertia to electromagnetism}

\author{Massimiliano Stengel}
\affiliation{Institut de Ci\`encia de Materials de Barcelona 
(ICMAB-CSIC), Campus UAB, 08193 Bellaterra, Spain}
\affiliation{ICREA - Instituci\'o Catalana de Recerca i Estudis Avan\c{c}ats, 08010 Barcelona, Spain}
\email{mstengel@icmab.es}

\author{Miquel Royo}
\affiliation{Institut de Ci\`encia de Materials de Barcelona 
(ICMAB-CSIC), Campus UAB, 08193 Bellaterra, Spain}
\email{mroyo@icmab.es}

\author{Emilio Artacho}
\affiliation{CIC Nanogune and DIPC, Tolosa Hiribidea 76, San Sebasti\'an, Spain}
\affiliation{Ikerbasque, Basque Foundation for Science, 48011 Bilbao, Spain}
\affiliation{Theory of Condensed Matter, Cavendish Laboratory, 
University of Cambridge, J. J. Thomson Avenue, Cambridge CB3 0HE, 
United Kingdom}

\date{\today}

\begin{abstract}
 {We establish a set of exact sum rules
that relate the interatomic force constants to the frequency-dependent 
electromagnetic susceptibility of {a solid or molecule}, thereby generalizing the 
long-established principles of rototranslational symmetry to the nonadiabatic regime.
Crucially, we show that in practical numerical implementations  
these sum rules are violated, unless special precautions are taken 
in the treatment of the atomic pseudopotentials.
We solve these issues once and for all by correctly adapting the pseudopotential 
to the motion of the corresponding nucleus, with a velocity dependence of
the nonlocal operator.
This prescription restores the correct Galilean covariance of the Schr\"odinger equation, and the expected identity
between mechanical rototranslations and electromagnetic perturbations.
These results conclusively fix a number of worrisome inconsistencies that were pointed out
over the years in the context of linear-response theory restoring, e.g., the validity of the Larmor 
theorem, and the equivalence between the inertial and electrical definitions of the Drude weight in metals.}
\end{abstract}

 %Lattice dynamics: first-principles theory
\maketitle

{\em Introduction.} {Translational and rotational symmetries play a central
role in the coupled dynamics of electrons and nuclei.~\cite{born/huang}
They ensure that physical laws remain constant regardless of position or orientation, and 
lead to conserved quantities (linear and angular momentum).
They also underpin a number of exact sum rules, relating microscopic interactions 
(e.g. the interatomic force constants or Born effective charges)
to macroscopic properties of the system.
Examples are numerous in both the physics and chemistry communities, ranging from
 the acoustic sum rule of Pick, Cohen and Martin~\cite{rmm_thesis}
and its recent generalization to metallic crystals~\cite{DreyerPRL22}, 
to the molecular sum rules involving the atomic axial~\cite{Buckingham-87}
or electron inertia~\cite{Scherrer-17} tensors. 
In addition to their obvious fundamental interest, such relations serve as stringent numerical tests for
practical computational methods, ensuring accuracy  
and avoiding spurious artefacts in the results.
In light of their importance, it is all the more surprising that
many of these sum rules are violated within the state-of-the-art 
implementations of density-functional theory, which typically rely on the pseudopotential approximation.}

Ionic pseudopotentials,~\cite{Bachelet-82,Hamann1979,troullier} either in the traditional Kleinman-Bylander~\cite{kleinman-82} form or in the more sophisticated modern constructions~\cite{vanderbilt:1990,bloechl:1994,Hamann-13}
are a mainstay of electronic-structure theory, and have played a decisive role in its long-lasting success.
Pseudopotentials allow for an efficient representation of the ionic cores 
with an accuracy that approaches that of all electron calculations, but at a comparably modest computational cost.
{Their application to nuclear dynamics, however, has revealed some crucial shortcomings as 
well.
Among the known issues, 
the violation of \emph{dynamical} rototranslational 
symmetries such as Galilean invariance and Larmor 
theorem~\cite{cyrus} 
appear especially worrisome, as it potentially affects all
situations where nuclei move in space with a nonzero 
velocity.
Such situations are ubiquitous in real-time dynamics, 
as well as in a broad range of physical properties 
going 
beyond the Born-Oppenheimer approximation.~\cite{Nafie-83,Scherrer-15}
These include, e.g., the atomic axial tensors (also known as dynamical magnetic moments~\cite{Zabalo-22}) entering the 
theory of vibrational spectroscopy~\cite{Stephens-85,Buckingham-87}, the electronic inertia tensor of Ref.~\cite{Scherrer-17}, or the 
dynamical Born charges in metals~\cite{DreyerPRL22}. 
Most sum rules involving this class of 
properties break down~\cite{Scherrer-17,DreyerPRL22,Zabalo-22} 
in presence of nonlocal pseudopotentials, reflecting the above 
issues.}

{These inconsistencies are likely to become all the more serious in the fully nonadiabatic regime.
The latter is attracting considerable 
interest in various area of condensed matter and quantum chemistry, ranging
from electron stopping power,~\cite{Zeb-12} ultrafast phenomena~\cite{Nisoli-17} and excited-state dynamics~\cite{Garrow-25} to others 
involving electron-phonon~\cite{SaittaPRL08,CalandraPRB10,GiustinoRMP17} and 
spin-phonon~\cite{RenPRX24,Royo-26} interactions.
Sum rules and exact relations like the ones above (which target the adiabatic limit),
should prove useful in all these contexts, but their computational study has been thwarted~\cite{nuria}
so far by the aforementioned difficulties with nonlocal operators.}
Most importantly, whether all these pitfalls reflect inherent limitations of the pseudopotential paradigm,
or an incorrect application thereof, remains unclear.

Here we show that the latter scenario applies. In particular, we demonstrate that many of the difficulties reported so far can be 
{resolved}
by introducing \emph{traveling pseudopotentials}, which ensure the correct propagation of the electronic wave functions in space and time. 
The central idea in this paper is simple: a pseudopotential that was
generated for a static nucleus at the origin has to be shifted in position
\textit{and in momentum} for the displaced, moving nucleus.
This prescription naturally restores both Galilean invariance and Larmor theorem and,
in combination with Pickard and Mauri's~\cite{Pickard-01,Pickard-03} treatment of uniform magnetic fields, 
leads to an exact correspondence between rototranslations and electromagnetic perturbations.
As a consequence, not only the undesirable discrepancy in the \emph{adiabatic-like} sum rules of Refs.~\cite{Scherrer-17,Zabalo-22,DreyerPRL22} disappears: 
we demonstrate, both analytically and numerically, how the kinetic inertia of the electron system under linear and angular accelerations matches its electrical and magnetic susceptibilities at any driving frequency.

The prescriptions presented here are uncomplicated to implement in modern time-dependent density-functional theory (DFT) and linear-response codes. 
They are immediately applicable to any type of pseudopotential, including PAW~\cite{bloechl:1994} and ultrasoft~\cite{vanderbilt:1990} ones.
In the following we illustrate the main concepts within a framework of 
Ehrenfest dynamics, with classical nuclei and quantum electrons treated within 
DFT.
It should be stressed, however, that the main ideas are much more general,
including explicitly correlated electrons, and/or quantum nuclei.
They even transcend pseudopotentials: all-electron calculations with
basis-set augmentation in core regions should similarly transform to
maintain Galilean invariance, {as done within Nafie's velocity-gauge formalism~\cite{Nafie-92}, 
or in traveling-orbital bases~\cite{Bates1953,Crothers1979,Kimura1985}.}

{\em General formalism.} We start from the Ehrenfest Lagrangian~\cite{Kwong1987,Todorov2001} 
\begin{equation}
\label{action}
\mathcal{L} =  i\hbar  \sum_n \langle \psi_n |\dot{\psi}_n \rangle
+\sum_{\kappa \alpha} \frac{M_\kappa}{2} \dot{R}^2_{\kappa \alpha} - E(\psi, \mathbf{R}_{\kappa} ), 
\end{equation}
which describes the coupled dynamics of a set of  {occupied single-particle orbitals} %electronic states 
$|\psi_n \rangle$ 
and of the $\mathbf{R}_{\kappa}$ positions of the classical nuclei of mass
$M_{\kappa}$ of the system. ($\alpha$ runs over the Cartesian coordinates, $\kappa$ is a sublattice index). 
The first two terms govern the time propagation of the active degrees of freedom,
while $E$ describes the potential energy. In the case of the Kohn-Sham energy functional,
the latter reads as
\begin{equation}
\label{e_ks}
\begin{split}
E_{\rm KS} = & E_{\rm kin} + E_{\rm loc}  {+ E_{\rm Ew}} {+ E_{\rm nl}},
\end{split}
\end{equation}
where $E_{\rm kin}$ refers to the electronic kinetic energy, $E_{\rm loc}$ contains all the contributions deriving from 
Hartree, exchange and correlation and the local part of the pseudopotential,  {and the Ewald energy $E_{\rm Ew}$ describes
the electrostatic ion-ion repulsion.}
Here we focus on the nonlocal pseudopotential term, $E_{\rm nl}=\langle \hat{V}^{\rm nl} \rangle$ (we indicate the
expectation value of a given operator $\hat{O}$ as $\langle \hat{O} \rangle = \sum_n \langle \psi_n |  \hat{O} | \psi_n \rangle$
henceforth). 
In static problems,  {the nonlocal operator is given by $\hat{V}^{\rm nl}= \sum_\kappa \hat{V}^{\rm nl}_\kappa$, where}
\begin{equation}
\label{vnl}
 {V}^{\rm nl}_\kappa({\bf r},{\bf r'}) = \sum_{l} {v}^{\rm nl}_\kappa({\bf r}-{\bf R}_{l\kappa},{\bf r'}-{\bf R}_{l\kappa}).
\end{equation}
(${\bf R}_{l\kappa}$
describes the atomic locations in the crystal, with 
$l$ indexing the cell.) 
Here we argue that the above formula must be revised whenever an atom is moving in space with a velocity $\dot{\bf R}_\kappa$.

{\em Traveling pseudopotentials.} The primary purpose of a pseudopotential consists in 
reproducing the properties of the all-electron (AE) atom as closely as possible within the target environment. 
Thus, before worrying about the actual pseudopotential construction, it is useful to clarify how the 
AE solution looks like for moving bodies. For an isolated atom traveling in space with 
a given velocity ${\bf v}$, and located at ${\bf r}=0$ at $t=0$, one has~{~\cite{Merzbacher}}
\begin{equation}
\label{travelpsi}
\psi'_{nlm}({\bf r},t) =  {e^{-i \Delta E t/ \hbar}} e^{i m_e {\bf v} \cdot {\bf r} / \hbar }\psi_{nlm}({\bf r}-{\bf v}t,t),
\end{equation}
where $\psi_{nlm}$ are solutions of the static Kohn-Sham equations for the AE atom, and $nlm$ refer to the 
principal and angular momentum quantum numbers.
It is easy to see that Eq.~\eqref{travelpsi} describes %\st{(modulo an energy phase)} 
a solution of the time-dependent Schr\"odinger equation for 
the traveling atom, with an excess kinetic energy of $\Delta E=m_e v^2 / 2$ and electronic linear momentum $\langle {\bf p} \rangle=m_e {\bf v}$; both $\Delta E$ and $\langle {\bf p} \rangle$ remain constant in time.

The usual \emph{Ansatz} of displacing the nonlocal pseudopotential rigidly in space via Eq.~\eqref{vnl} manifestly fails at
reproducing this result, and is therefore unfit to dealing with time-dependent problems, \emph{even within the adiabatic approximation}.
This issue, however, is fixed once and for all by introducing a velocity dependence in the pseudopotential as follows
(we assume Hartree atomic units, with $m_e=\hbar=1$, from now on) 
\begin{equation}
\label{travelv}
 {V}^{\rm nl}({\bf r},{\bf r'}) = \sum_{l \kappa}  {e^{i {\dot{{\bf R}}_{\kappa}} \cdot ({\bf r-r'})} } {v}^{\rm nl}_\kappa({\bf r}-{\bf R}_{l\kappa},{\bf r'}-{\bf R}_{l\kappa}).
\end{equation}
This prescription, straightforward to implement in a plane-wave pseudopotential code,~\footnote{
For a periodic crystal, it boils down to operating a shift in momentum space as ${\bf k} \rightarrow {\bf k} - {m_e  \dot{\bf R}_{l\kappa}}/{\hbar}$.
}
ensures that the pseudo wave functions transform according to Eq.~\eqref{travelpsi} 
under a Galilean boost (see Ref.~\cite{supp}, Section I),
and exactly reproduce the AE results for 
the energy and linear momentum. \nocite{gonze/lee,Baroni-01,SchiaffinoPRB19,Resta-25,pseudodojo,Beigi-01}
 {(A generalization to PAW~\cite{bloechl:1994} and localized-orbital basis sets is provided in Sec.~II.C of~\cite{supp}.)}
Note that Galilean covariance now holds~\cite{supp}
for the microscopic electrical current as well: the latter correctly transforms as  
${\bf j}'({\bf r}) = {\bf j}({\bf r}-{\bf v}t) + {\bf v} \rho({\bf r}-{\bf v}t)$,
where $\rho({\bf r},t)$ is the total charge density.

{\em Energy and linear momentum.} 
By applying Noether's theorem to infinitesimal shifts in time and space of the Ehrenfest Lagrangian, 
Eq.~\eqref{action}, we obtain the conserved energy ($K$) and total momentum ($P_\alpha$), respectively, as
\begin{align}
K =& \sum_{\kappa \alpha} \left( \frac{M_\kappa}{2} \dot{R}^2_{\kappa \alpha} +  A^{\rm nl}_{\kappa\alpha} \dot{R}_{\kappa\alpha} \right)
  + E_{\rm KS}, \\
P_\alpha =& \sum_{\kappa} \left( M_\kappa \dot{R}_{\kappa \alpha} + A^{\rm nl}_{\kappa\alpha} \right)
  + \langle \hat{p}_\alpha \rangle.
\end{align}
%
%(We drop the Bravais index $l$ and assume cell-periodic dynamics henceforth.)
The only difference with respect to the traditional definitions consists in the appearance of the nonlocal gauge field,
$A^{\rm nl}_{\kappa\alpha} = -\left\langle {\partial \hat{V}^{\rm nl} } / { \partial \dot{R}_{\kappa \alpha} } \right\rangle$.
After observing that $\sum_\kappa A^{\rm nl}_{\kappa\alpha} = i
 \langle [ \hat{V}^{\rm nl}, r_\alpha] \rangle$ \cite{supp}, 
we can write the conserved total momentum as
\begin{equation}
\label{ptot}
P_\alpha = \sum_{\kappa}  M_\kappa \dot{R}_{\kappa \alpha} 
  + i  \langle [H,r_\alpha] \rangle,
\end{equation}
where  {$H$ is the instantaneous Hamiltonian}. The electronic contribution is the expectation value of the velocity operator,
which describes the time derivative of the position following Ehrenfest's theorem, and hence
the physical particle current flowing through the system at time $t$.
{Most importantly, the electronic contribution to the total mechanical momentum coincides 
(modulo the expected prefactor of $-e/m_e$) with the established definition of the macroscopic charge current 
in electromagnetism. 
This is a central result, as it restores the exact equivalence between the effects of mechanical acceleration and
electric fields, characterizing any physical particle endowed with a nonzero mass and charge.~\cite{StengelPRB18}.

{{\em Angular momentum.} In isolated systems, rotational invariance leads to a third set of constants of motion, 
in the form of a conserved total angular momentum
\begin{equation}
\label{ltot}
\begin{split}
{\bf L} =&  \sum_{\kappa}  M_\kappa {\bf R}_\kappa \times \dot{\bf R}_{\kappa}
  + \langle \hat{\bf L} \rangle +  \langle \hat{\bf S} \rangle,
\end{split}  
\end{equation}
where  the electronic orbital angular momentum operator now becomes,
\begin{equation}
\label{lorb}
\hat{\bf L}  = {\bf r} \times {\bf p} + \sum_{\kappa}  {\bf R}_{\kappa} \times i \left[ V^{\rm nl}_\kappa, {\bf r}  \right],
\end{equation}
and $\hat{S}_\alpha=\hbar \sigma_\alpha/2$ describes the spin contribution.
Remarkably, Eq.~\eqref{lorb} coincides, modulo a trivial unit conversion prefactor, with the first-order Hamiltonian in a uniform magnetic field established by Pickard and Mauri~\cite{Pickard-03}.
This result proves that our prescription not only yields the physically correct behavior of the Hamiltonian under Galilean boosts, but also under uniform rotations. 
By doing so, it solves a long-standing problem, in that the traditional treatment of nonlocal pseudopotentials leads to a problematic violation of the Larmor theorem~\cite{cyrus}. 
Our formalism, in combination with that of Ref.~\cite{Pickard-03}, restores the exact gyromagnetic ratio between the angular momentum and the magnetic moment, $\langle \hat{\bf L} \rangle/ 2c = \langle \partial \hat{H} /\partial {\bf B} \rangle $, thereby enabling the calculation of properties that depend on the orbital magnetization response (e.g., the rotational $g$-factor of molecules) by purely mechanical means.
}

\emph{Adiabatic regime.} As an additional remark, it is insightful to note that $A^{\rm nl}_{\kappa \alpha}$ enter as a nonlocal correction to the Mead-Truhlar 
gauge fields~\cite{Mead-79}, playing the role of the Berry connection $A_{\kappa\alpha}$ in configuration space~\cite{David-book} within Born-Oppenheimer dynamics. 
We find~\cite{supp}
 \begin{equation}
\label{berry}
A_{\kappa\alpha} = i \sum_n \Big\langle \psi_n \Big| \frac{\partial \psi_n}{\partial R_{\kappa \alpha}} \Big\rangle + A^{\rm nl}_{\kappa \alpha},
\end{equation}
where the first term on the rhs is the standard Berry connection.
This is another important result: it tells us that traveling nonlocal pseudopotentials entail a finite contribution to 
the total Berry phase \emph{even if the system evolves infinitely slowly along a given structural path}. Such a correction is sufficient 
to guarantee that the conserved total angular momentum~\cite{Bian-23} within the adiabatic regime is correctly given by Eqs.~\eqref{ltot} and~\eqref{lorb}.

{\em Linear response.} 
We now 
consider a time-dependent perturbation of the crystal structure in the form 
\begin{equation}
\label{phonon}
R_{\kappa \alpha}(t) = R^0_{\kappa \alpha} + u_{\kappa \alpha}^\w e^{-i\omega t},
\end{equation}
where ${\bf R}^0_\kappa$ are the atomic positions at rest. 
The second derivatives of the action, $\mathcal{A} = 1/T \int_0^T dt \mathcal{L}$,
with respect to Eq.~\eqref{phonon} are related to the zone-center dynamical 
force constants (dynFC), ${\bf C}(\w)$, via~\cite{Royo-26}
\begin{equation}
M_{\kappa\alpha,\kappa'\beta} \w^2 - C_{\kappa\alpha,\kappa'\beta}(\w) = \frac{\partial^2 \mathcal{A}}{\partial u^{-\w}_{\kappa\alpha} \partial u^{\w}_{\kappa'\beta}},
\end{equation}
where $M_{\kappa\alpha,\kappa'\beta} = M_{\kappa} \delta_{\kappa \kappa'} \delta_{\alpha\beta}$ is the bare nuclear mass tensor.
 {The dynFC are central quantities in the theory of electron-phonon couplings~\cite{GiustinoRMP17}, and 
a key ingredient when calculating nonadiabatic effects in the phonon~\cite{SaittaPRL08,CalandraPRB10} and coupled
spin-phonon sectors~\cite{RenPRX24,Royo-26}.}
Similarly, we consider a uniform electric field oscillating in time as
$\bm{\mathcal{E}}(t) = \bm{\mathcal{E}}^\w e^{-i\omega t}.$
The corresponding second derivatives yield the established definition of the dielectric susceptibility tensor ($\Omega$ is the cell volume),
\begin{equation}
\chi_{\alpha\beta}(\w) = -\frac{1}{\Omega}\frac{\partial^2 \mathcal{A}}{\partial \mathcal{E}^{-\w}_{\alpha} \partial \mathcal{E}^{\w}_{\beta}},
\end{equation}
which relates to the macroscopic dielectric tensor via $\epsilon_{\alpha\beta}(\w) = \delta_{\alpha\beta} + 4\pi \chi_{\alpha\beta}(\w)$.
Finally, the mixed derivatives yield the Born effective charges, $Z_{\kappa\alpha,\beta}(\w)=Z_\kappa \delta_{\alpha\beta} + Z^{\rm el}_{\kappa\alpha,\beta}(\w)$,
which depend as usual on the bare pseudopotential charge $Z_\kappa$ and on an additional electronic contribution
$Z^{\rm el}_{\kappa\alpha,\beta}(\w)$.

{\em Translational sum rules.} By incorporating our traveling pseudopotentials~\cite{supp} into the aforementioned definitions of 
the force constants, ${\bf C}(\w)$, and Born effective charges, ${\bf Z}(\w)$,
we obtain the following exact sum rules, valid at any frequency for insulators and metals alike,
\begin{subequations}
\label{sum_tra}
\begin{align}
\sum_{\kappa'} C_{\kappa \alpha,\kappa' \beta}(\w) =&  \w^2  Z^{\rm el}_{\kappa\alpha,\beta}(\w), \label{sum_ck}\\
\sum_{\kappa} Z_{\kappa \alpha,\beta}(\w) =&  
 -\w^2 \Omega \chi_{\alpha\beta}(\w). \label{bec} \\
\frac{1}{\w^2} \sum_{\kappa \kappa'} C_{\kappa \alpha,\kappa' \beta}(\w)  =& - N \delta_{\alpha \beta}-\w^2 \Omega \chi_{\alpha\beta}(\w). \label{sum_ckk}
\end{align}
\end{subequations}
{($N$ is the total number of valence electrons per cell.)}
Eq.~\eqref{sum_ck} states that an acceleration of the whole lattice, $\dot{\bf v}$, results in atomic forces
that are indistinguishable from those produced by a uniform electric field, $\bm{\mathcal{E}}$, applied at the same frequency $\w$.
Eq.~\eqref{bec} reflects the same equivalence principle by focusing, this time, on the identity between
the macroscopic polarization that results from either $\dot{\bf v}$ or $\bm{\mathcal{E}}$.
Eq.~\eqref{sum_ckk} follows directly from the combination of Eq.~\eqref{sum_ck}  and Eq.~\eqref{bec}.
 {As a whole, Eq.~\eqref{sum_tra} generalize the classic results of Pick, Cohen and Martin~\cite{rmm_thesis} 
to the nonadiabatic case.}

To illustrate the physical significance of Eq.~\eqref{sum_ckk}, 
it is useful to introduce the electronic inertia tensor, ${\bf M}^{\rm el}$, via the 
following adiabatic expansion,
\begin{equation}
\label{mass}
{\bf C}(\w) = {\bf C}(0) - \w^2 {\bf M}^{\rm el} + O(\w^3).
\end{equation}
The latter describes the mechanical inertia of the electrons that follow adiabatically 
the trajectory of the system in configuration space.~\cite{Scherrer-17,Royo-26}
Then, assuming that $\chi$ is regular in the static limit, Eq.~\eqref{sum_ckk} can be expressed as
$\sum_{\kappa \kappa'} M^{\rm el}_{\kappa \alpha \kappa' \beta} = N \delta_{\alpha\beta}$.
This means that, in an insulator, the excess inertial mass associated to a rigid translation of the 
lattice corresponds to exactly $Nm_e$, consistent with the 
Galilean covariance of our theory. {This result can be regarded as a manifestation of the $f$-sum 
rule in the kinematic context~\cite{Goldhaber-05}, and reduces to Eq.~(17) of Scherrer {\em et al.}~\cite{Scherrer-17}
in the special case of a local Hamiltonian. The present formalism generalizes its validity to the
nonlocal pseudopotential framework, setting the stage for accurate first-principles calculations
of electron inertia effects in a broad variety of systems.}

{\em Metals.} Eq.~\eqref{bec} was first derived in Ref.~\onlinecite{DreyerPRL22}, and applied to metals.~\footnote{
 Since the susceptibility is related to the conductivity via $\bm{\chi} = i\bm{\sigma}/\w$, 
 the rhs of Eq.~\eqref{bec} can be equivalently written as $-i\w \sigma_{\alpha\beta}(\w)$.
 This indicates that a minus sign is missing in Eq.~(5) of Ref.~\onlinecite{DreyerPRL22}.}
 There the electrical susceptibility 
diverges in the low-frequency limit as
\begin{equation}
\label{chi_drude}
\chi_{\alpha\beta}(\w) = -\frac{D_{\alpha\beta}}{\pi \w^2} + \cdots,
\end{equation}
where $D_{\alpha\beta}$ is the Drude weight tensor, a defining geometric property of the metallic state.
Thus, by combining Eq.~\eqref{chi_drude} and Eq.~\eqref{bec} one obtains~\cite{DreyerPRL22}
\begin{equation}
\frac{1}{\Omega} \sum_{\kappa} Z_{\kappa \alpha,\beta}(\w \rightarrow 0) = \frac{D_{\alpha\beta}}{\pi}.
\end{equation}
This sum rule is, however, violated if the propagation of the 
atomic pseudopotentials is treated in the usual way~\cite{DreyerPRL22}.
This undesirable inconsistency of the existing 
linear-response theory, which has remained unresolved to date, is fixed once
and for all once the correct velocity dependence of the nonlocal operator is
accounted for via Eq.~\eqref{travelv}.
We regard this as a major formal achievement of this work.

Meanwhile, our formalism corroborates 
an alternative physical interpretation of the Drude weight that 
was proposed recently,~\cite{nuria}
as the leftover kinetic energy in a nonadiabatic boost of the crystal.
To see this explicitly, we use Eq.~\eqref{sum_ck} in combination with
Eq.~\eqref{bec} and Eq.~\eqref{mass} to write the total inertial mass of the electrons as
\begin{equation}
\label{sum_ckk2}
\begin{split}
\sum_{\kappa \kappa'} M^{\rm el}_{\kappa \alpha,\kappa' \beta}  =&  N \delta_{\alpha \beta} -  \frac{\Omega D_{\alpha\beta}}{\pi},
\end{split}
\end{equation}
which matches the conclusions of Ref.~\cite{nuria}.
The present formalism guarantees that Eq.~\eqref{sum_ckk2} holds exactly even in presence of nonlocal 
pseudopotentials, thereby resolving the technical difficulties that were pointed out
in Ref.~\cite{nuria}; this is another central result of this work.

\begin{table}
\setlength{\tabcolsep}{3pt}
\begin{center}
\caption{\label{eq20bc}  {Numerical test of} the sum rules in Eqs.~\eqref{sum_tra} for bulk crystals. 
Values in parentheses are obtained by neglecting 
the velocity dependence of $\hat{V}^{\rm nl}$ in Eq.~\eqref{travelv}. 
The first column serves as a reference, while the second and third columns are reported as relative and absolute deviations, respectively. Values in the third column should correspond to the number of valence electrons: $N_{\rm GaP}=18$, $N_{\rm SrTiO_3}=40$ and $N_{\rm C}=8$.
We set $\hbar\omega=0.04$ Ha 
in all cases.}
\begin{tabular}{c|ccc}\hline\hline
   \T\B & \multicolumn{1}{c}{$-\omega^2\Omega\chi_{\alpha\alpha}(\omega)$} & \multicolumn{1}{c}{$\sum\limits_{\kappa} Z_{\kappa\alpha,\alpha}(\omega)$} &
   \multicolumn{1}{c}{$\frac{1}{\omega^2}\sum\limits_{\kappa,\kappa'} C_{\kappa\alpha,\kappa'\alpha}(\omega)$}  \\
   \hline
   \T\B GaP       & $-$0.3507 & $<$0.01 (1.04)  & 18.000 (23.97) \\
   \T\B SrTiO$_3$ & $-$0.2701 & $<$0.01 (6.33)  & 39.998 (45.12) \\
   \T\B C         & $-$0.0452 &    0.01 (5.25)  & 8.000 (8.48) \\
   \hline \hline
\end{tabular}
\end{center}
\end{table}

{\em Rotational sum rules.} In isolated systems, {the exact correspondence between magnetic fields and 
angular velocities leads to the following rotational sum rules for the electron inertia tensor,
%\begin{align}
% \sum_{\kappa \gamma\lambda } 
% \epsilon^{\alpha \gamma \lambda} {R}_{\kappa \gamma}
 % {M}^{\rm el}_{\kappa \lambda, \kappa' \beta} = &
% -2 \frac{\partial m^{\rm el}_\alpha }{\partial \dot{R}_{\kappa' \beta}}.
%\end{align}
\begin{subequations}
\label{sum_rot}
\begin{align}
  \left[ (\hat{\alpha} \times {\bf R}) \cdot  {\bf M}^{\rm el} \right]_{\kappa \beta} =& \frac{\partial \langle \hat{L}_\alpha \rangle}{\partial \dot{R}_{\kappa \beta}} = 
% \epsilon^{\alpha \gamma \lambda} {R}_{\kappa \gamma}
 % {M}^{\rm el}_{\kappa \lambda, \kappa' \beta} = &
 -2 \frac{\partial m^{\rm el}_\alpha }{\partial \dot{R}_{\kappa \beta}}, \label{axial} \\
%\end{align}
%
%
%[...] it is natural to expect the excess mass to affect the rotational inertia of the system as well. 
%Indeed, our formalism leads to the following sum rule,
%\begin{align}
(\hat{\alpha} \times {\bf R}) \cdot {\bf M}^{\rm el} \cdot (\hat{\beta} \times {\bf R}) =& \frac{\partial \langle \hat{L}_\alpha \rangle}{\partial \dot{\theta}_{\beta}} = % I^{\rm el}_{\alpha \beta},
4 \chi^{\rm para}_{\alpha \beta}.
 \label{sum_l2} 
\end{align}
\end{subequations}
}
Here ${\bf R}$ stands for the atomic positions at rest of all atoms (the scalar products involve a double sum over the sublattice index and Cartesian directions), $\hat{\bf L}$ is given by Eq.~\eqref{lorb}, and $\dot{\bm{\theta}}$ is the angular velocity {pseudovector}.
{To the best of our knowledge, neither Eq.~\eqref{axial} nor Eq.~\eqref{sum_l2} were pointed out before.}

{Eq.~\eqref{axial} links ${\bf M}^{\rm el}$ %relates the first moment of the inertia tensor 
to the derivative of the electronic angular momentum with respect to the atomic velocity. The latter, in turn, corresponds (modulo a prefactor of $-2$) to the 
electronic part of the atomic axial tensors,~\cite{Stephens-85,Buckingham-87} ${\partial m_\alpha }/{\partial \dot{R}_{\kappa \beta}}$. 
  (${\bf m} = {\bf m}^{\rm el} + {\bf m}^{\rm ion}$
  is the total magnetic moment, inclusive of the ionic point charge contribution ${\bf m}^{\rm ion} = \frac{1}{2}\sum_\kappa Z_\kappa {\bf R}_\kappa \times \dot{\bf R}_\kappa$.)
Eq.~\eqref{sum_l2} relates the second moment of ${\bf M}^{\rm el}$ to the derivative of the electronic angular momentum with respect to the angular velocity.}
By definition, ${I^{\rm el}_{\alpha \beta} =} {\partial \langle L_\alpha \rangle}/{\partial \dot{\theta}_{\beta}}$
describes a purely electronic contribution to the overall moment of inertia, and relates to the paramagnetic susceptibility tensor %~\cite{rotchi_note}
via $\chi^{\rm para}_{\alpha \beta}=I^{\rm el}_{\alpha \beta}/4$.
${\bf I}^{\rm el}$ also bears a direct relation to the 
geometric orbital magnetization~\cite{Trifunovic-19,Zabalo-22} at first order 
in the angular velocity, %and hence to the rotational $g$-factor,
\begin{equation}
\label{dmdw}
\frac{\partial  m_\alpha }{\partial \dot{\theta}_{\beta}} = -\frac{1}{2} \left[ I^{\rm el}_{\alpha\beta} - (\hat{\alpha} \times {\bf R}) \cdot {\bf Z} \cdot (\hat{\beta} \times {\bf R}) \right],
\end{equation}
($Z_{\kappa\alpha,\kappa'\beta}= \delta_{\kappa\kappa'} \delta_{\alpha\beta} Z_\kappa$ is the bare pseudopotential charge tensor) and hence to the rotational $g$ factor~\cite{Ceresoli-02,Zabalo-22}, via $g = (2/I_{\alpha\alpha}) {\partial  m_\alpha }/{\partial \dot{\theta}_{\alpha}}$. 
{By plugging Eq.~\eqref{axial} into Eq.~\eqref{sum_l2}, one 
recovers the magnetizability sum rule of Buckingham {\em et al.}~\cite{Buckingham-87},
\begin{equation}
\label{magsum}
\sum_\kappa \epsilon^{\beta \gamma \lambda} R_{\kappa \gamma} \frac{\partial m^{\rm el}_\alpha }{\partial \dot{R}_{\kappa \lambda}} = -2 \chi^{\rm para}_{\alpha \beta},
\end{equation}
which corroborates the internal consistency of our findings.}
The combination of our traveling pseudopotentials with Pickard and Mauri's (PM) formulation of the uniform magnetic field perturbation guarantees that
all the above relations are exactly fulfilled in presence of nonlocal operators, thereby overcoming the long-standing difficulties of the existing theory.~\cite{Zabalo-22}

 {Eq.~\eqref{sum_tra} %,~\eqref{dmdw} 
 and~\eqref{sum_rot} constitute our main formal results from the point of view of the physics.}
To demonstrate them numerically, 
we have implemented our prescriptions in the linear-response module of ABINIT~\cite{abinit}.
In Table~\ref{eq20bc} we present a numerical test of the translational sum rules on three representative crystals.
By accounting for the velocity dependence of the pseudopotentials, the sum rules are satisfied to a very high accuracy, correcting for the large discrepancies that arise within the standard implementation. Such deviations are especially large in the case of the force constants, where the effective number of electrons contributing to inertia departs significantly (10\%--20\%) from $N$ in all cases.
In Table~\ref{tab_magnetic}, we test the rotational sum rules by extracting the rotational $g$ factor and magnetic susceptibility ($\chi^{\rm mag}_{\alpha\alpha}$) of a set of small molecules. 
We choose our target system because of the availability of benchmark all-electron data for $\chi^{\rm mag}_{\alpha\alpha}$, which were calculated in Ref.~\cite{Pickard-03} within the individual gauges for atoms in molecules (IGAIM) method.~\cite{igaim}
Our results are in remarkable agreement with the IGAIM values, with maximum deviations of $0.1 \times 10^{-6}$ cm$^3$/mol (0.5\%) and 
$0.06 \times 10^{-6}$ cm$^3$/mol (0.3\%) in the case of CH$_4$ and P$_2$, respectively.
Such an excellent match provides strong support to our arguments, and to the overall numerical quality of our implementation.
Regarding $g$ factors, our values are generally in line with those of Ref.~\cite{Zabalo-22}, with more important deviations in some cases~\cite{supp}; a detailed account of these results will be presented elsewhere.

\begin{table}
\setlength{\tabcolsep}{6pt}
\begin{center}
\caption{\label{tab_magnetic}  
Magnetic susceptibilities {$\bm{\chi}^{\rm mag} = \bm{\chi}^{\rm para} + \bm{\chi}^{\rm dia}$} (in units of $10^{-6}$ cm$^3$/mol) and rotational $g$ factors calculated via Eqs.~\eqref{sum_l2} and \eqref{dmdw}. The rigid-core contribution, which we calculate via an atomic all-electron code, is accounted for in the diamagnetic term, {$\chi^{\rm dia}_{\alpha \beta} = -\frac{1}{4}  \int d^3 r 
\left( r^2 \delta_{\alpha \beta} - r_\alpha r_\beta \right)
n^{(0)}({\bf r})$, where $n^{(0)}({\bf r})$ is the ground-state electron density.}}
\begin{tabular}{r|rrrr}\hline\hline
   \T\B & \multicolumn{1}{c}{$g$ factor} & \multicolumn{1}{c}{$\mathcal{\chi}^{\rm mag}_{xx}$} &
   \multicolumn{1}{c}{$\mathcal{\chi}^{\rm mag}_{yy}$} & \multicolumn{1}{c}{$\mathcal{\chi}^{\rm mag}_{zz}$}  \\
   \hline
%   \T\B  H$_2$  &    0.8829  & $-$3.8950 & $-$4.4517  & $-$4.4517 \\
%   \T\B  N$_2$  & $-$0.2954  & $-$18.1028 & $-$8.9519  & $-$8.9519 \\
%   \T\B  F$_2$  & $-$0.1264  & $-$17.4679 & $-$6.8350  & $-$6.8350\\
   \T\B  P$_2$  & $-$0.1165  & $-$46.12 & $-$18.16 & $-$18.16 \\
%   \T\B  HF$_2$ &    0.7419  & $-$10.6120  & $-$17.2100 & $-$17.2100\\
   \T\B  CH$_4$ &    0.3526  & $-$19.97 & $-$19.97 & $-$19.97 \\
 %  \T\B  C$_6$H$_6$  &   $-$0.0812      &      $-$33.33      &   $-$33.33       & $-$99.68 \\
    \T\B  C$_6$H$_6$  &   $-$0.0811      &      $-$33.51      &   $-$33.51       & $-$98.93 \\
     \hline \hline
\end{tabular}
\end{center}
\end{table}

In summary, we have discussed a general strategy to achieve all-electron accuracy in dynamical problems where the nuclei are represented by nonlocal pseudopotentials. 
Given its simplicity and effectiveness, we expect it to have considerable impact on all areas of research involving first-principles simulations with nuclear dynamics. 
Within the context of linear-response theory, it will be interesting to assess the importance of these corrections on
the calculation of spatial dispersion (e.g., flexoelectricity~\cite{Royo2019},  {vibrational circular dichroism~\cite{Nafie-92,Scherrer-15,Ditler-22}}, etc.), and time-dispersion (e.g., molecular Berry curvatures~\cite{RenPRX24})
properties.
Our results, in combination with the recently developed tools~\cite{RenPRX24,Royo-26} to treat the coupled 
spin-phonon problem, open the door to an exact computational realization of angular momentum coupling effects 
(e.g., Einstein-de-Haas). 
We regard all these as very promising avenues for future studies.

\begin{acknowledgments}
This work was supported by the Spanish MCIN/AEI/10.13039/501100011033 
through grants PID2019-107338RB-C61, PID2022-139776NB-C65, 
PRX22/00390, and PID2023-152710NB-I00, and a Mar\'{\i}a de 
Maeztu award to Nanogune, Grant CEX2020-001038-M 
and a Severo Ochoa Excellence award to ICMAB, Grant
CEX2023-001263-S, as well as by the Generalitat de Catalunya
through Grant No. 2021 SGR 01519, and by 
the United Kingdom's EPSRC Grant no. EP/V062654/1.
We are grateful to Francesco Mauri, Raffaele Resta, Nuria Santerv\'as-Arranz, Ivo Souza and David Vanderbilt for many stimulating discussions.
\end{acknowledgments}

\bibliography{merged}

\end{document}

%\title{Supplemental material for \\ ``Pseudopotentials that move: From inertia to electromagnetism''}
\title{Supplemental material for \\  {``Rototranslational sum rules for %nonadiabatic lattice 
{nuclear} dynamics via traveling pseudopotentials''}}

\author{Massimiliano Stengel}
\affiliation{Institut de Ci\`encia de Materials de Barcelona 
(ICMAB-CSIC), Campus UAB, 08193 Bellaterra, Spain}
\affiliation{ICREA - Instituci\'o Catalana de Recerca i Estudis Avan\c{c}ats, 08010 Barcelona, Spain}
\email{mstengel@icmab.es}

\author{Miquel Royo}
\affiliation{Institut de Ci\`encia de Materials de Barcelona 
(ICMAB-CSIC), Campus UAB, 08193 Bellaterra, Spain}
\email{mroyo@icmab.es}

\author{Emilio Artacho}
\affiliation{CIC Nanogune and DIPC, Tolosa Hiribidea 76, San Sebasti\'an, Spain}
\affiliation{Ikerbasque, Basque Foundation for Science, 48011 Bilbao, Spain}
\affiliation{Theory of Condensed Matter, Cavendish Laboratory, 
University of Cambridge, J. J. Thomson Avenue, Cambridge CB3 0HE, 
United Kingdom}

\date{\today}

\maketitle

\section{Introduction}

In these notes we spell out some technicalities in support of the results presented in the main text. We also provide 
additional data to complement the numerical tests therein.

\section{Galilean invariance}

\subsection{Wave functions}

The time-dependent Schr\"odinger equation (TDSE) reads as
\begin{equation}
i {\hbar} | \dot{\psi}_{n} \rangle = \hat{H}(t)|{\psi}_{n} \rangle,
\end{equation}
where $\hat{H}(t) = \hat{T} + \hat{V}(t)$
We use the following transformation for the wave functions and atomic coordinates,
\begin{align} 
\psi'_{n} ({\bf r}, t) = & e^{im_e {\bf v\cdot r}/\hbar - i\Delta E t/\hbar} {\psi}_{n} ({\bf r}-{\bf v}t, t), \\
{{\bf R}'_\kappa(t)} =& %
{{\bf R}_\kappa(t) {+} {\bf v}t,}
\end{align}
We now consider the transformed TDSE,
\begin{equation}
i {\hbar}  | \dot{\psi}'_{n} \rangle = \hat{H}'(t)|{\psi}'_{n} \rangle.
\end{equation}
We have, at any given time  {(using $\hbar=m_e=1$ from now on)}, 
\begin{subequations}
\label{galilei}
\begin{align}
i|\dot{\psi}'_{n} \rangle \doteq& e^{i {\bf v\cdot r} - i {\Delta E} t} \left( i  |\dot{\psi}_{n} \rangle +\hat{\bf p} \cdot {\bf v} |{\psi}_{n} \rangle  +  \Delta E |{\psi}_{n} \rangle \right), \\
\hat{T}  |{\psi}'_{n} \rangle \doteq & e^{i {\bf v\cdot r}- i {\Delta E} t} \left( \hat{T}  + 
 \hat{\bf p} \cdot {\bf v} + \frac{m}{2}v^2 \right) |{\psi}_{n} \rangle, \\
%\langle {\bf r} |
\hat{V}' |{\psi}'_{n} \rangle \doteq & e^{i {\bf v\cdot r} - i {\Delta E} t} %\langle {\bf r} - {\bf v}t |
\hat{V} |{\psi}_{n} \rangle. \label{transv}
\end{align}
\end{subequations}
{where the dotted equal sign between a primed and unprimed ket, $|v'\rangle \doteq |v\rangle$,
stands for $\langle {\bf r} |v'\rangle = \langle {\bf r} - {\bf v}t |v\rangle$.}
Thus, we find that the boosted states are solution of the boosted TDSE, provided that we set 
\begin{equation}
\Delta E =  \frac{m_e}{2}v^2.
\end{equation}

 {Our prescription for the propagation of the nonlocal operators is essential to ensure that Eq.~\eqref{galilei} holds; otherwise,
the transformed wavefunctions are no longer solutions of the boosted TDSE, and Galilean covariance breaks down.
To see this, consider the nonlocal pseudopotential operator applied to the boosted wave functions.
Within the standard prescriptions, Eq.(3) of the main text, one has
\begin{equation}
\begin{split}
%\langle {\bf r}|
 \hat{V}^{\rm nl'} |\psi'_n \rangle \doteq & e^{- i {\Delta E} t} %\langle {\bf r} - {\bf v}t |
 \hat{V}^{\rm nl} e^{i {\bf v\cdot r}} |{\psi}_{n} \rangle \\
 \neq & e^{i {\bf v\cdot r} - i {\Delta E} t} %\langle {\bf r} - {\bf v}t |
 \hat{V}^{\rm nl} |{\psi}_{n} \rangle.
\end{split}
\end{equation}
Crucially, the second equality is violated because the nonlocal operator does \emph{not} commute with the position operator, 
\begin{equation}
[\hat{V}^{\rm nl}, {\bf r}] \neq 0. 
\end{equation}
As a consequence, Galilean invariance is violated as well, since it requires Eq.~\eqref{transv}
to hold.
Conversely, once the velocity dependence of the pseudopotential is correctly
accounted for via Eq.(5) of the main {text}, the validity of Eq.~\eqref{transv} is restored,
\begin{equation}
\begin{split}
%\langle {\bf r}| 
\hat{V}^{\rm nl'} |\psi'_n \rangle \doteq& e^{- i {\Delta E} t} %\langle {\bf r} - {\bf v}t | 
e^{i {\bf v\cdot r}} \hat{V}^{\rm nl} e^{-i {\bf v\cdot r}} e^{i {\bf v\cdot r}} |{\psi}_{n} \rangle \\
 \doteq & e^{i {\bf v\cdot r} - i {\Delta E} t} 
 %\langle {\bf r} - {\bf v}t |
 \hat{V}^{\rm nl} |{\psi}_{n} \rangle.
\end{split}
\end{equation}
}

%\subsection{Band energies}

 {Before closing this section, we briefly comment on the effects of a Galileo transformation 
on the electronic structure. 
To this end, we calculate the expectation value of the boosted Hamiltonian on 
the $n$-th wave function.
From Eqs.~\eqref{galilei}, we find 
\begin{equation}
\label{banden}
\epsilon'_n = \langle \psi'_n | \hat{H}'|\psi'_n \rangle = \epsilon_n + \langle {\bf p} \rangle_n \cdot {\bf v} + \frac{m}{2}v^2.
\end{equation}
If we assume an isolated system at rest in its electronic ground state, the 
expectation value of the momentum operator vanishes for all $n$: $\langle {\bf p} \rangle_n=0$.
Then, in the boosted system all band energies are shifted upwards by the same amount, which 
means that their relative difference is the same as in the system at rest.
Following the discussion of the previous paragraph, the proper use
of traveling pseudopotentials in the boosted system is key for this result to hold.}

\subsection{Current density}

The microscopic current density is given as the expectation value of the current density operator.
Following Ref.~\cite{cyrus} we define the latter as minus the derivative of the instantaneous Hamiltonian 
with respect to the vector potential field.
It consists of a local part and a nonlocal contribution,
\begin{equation}
\hat{J}_\alpha = \hat{J}^{\rm loc}_\alpha + \hat{J}^{\rm nl}_\alpha,
\end{equation}
which are respectively given by
\begin{align}
\hat{J}^{\rm loc}_\alpha({\bf r}) =& -\frac{ |{\bf r} \rangle \langle {\bf r}| \hat{p}_\alpha +  \hat{p}_\alpha |{\bf r} \rangle \langle {\bf r}|}{2},\\
\hat{J}^{\rm nl}_\alpha({\bf r}) =&  {\int d^3 s \int d^3 s' |{\bf s} \rangle \langle {\bf s}' | \times} \nonumber \\ 
& \sum_\kappa e^{i\dot{\bf R}_\kappa \cdot ({\bf s-s'})} V^{\rm nl}_\kappa ({\bf s,s'}) \phi_\alpha^{\bf (r)}({\bf s,s'}).
\end{align}
Here the factor $\phi_\alpha^{\bf (r)}$ is defined as follows,
\begin{equation}
 \phi_\alpha^{\bf (r)}({\bf s,s'}) = i \frac{\delta }{\delta A_\alpha({\bf r})}\int_{\bf s'}^{\bf s} {\bf A}\cdot d \bm{\ell} .
 \end{equation}
 {If ${\bf A}$ is uniform in space, this formula correctly yields $ \phi_\alpha = i (s_\alpha - s'_\alpha)$, recovering the usual formula for the 
macroscopic current in presence of nonlocal potentials, $\hat{J}_\alpha = -i[\hat{H},{\bf r}]$.}
In the most general case, $\phi_\alpha$ is an imaginary function of the three real variables ${\bf r}$, ${\bf s}$ and ${\bf s}'$. Generally, the path of integration needs to be specified, as the results depend on it. Nevertheless, in the present context there is no need to worry about it, as the current density transforms correctly regardless of the assumed path.

To see this, we can apply the above definitions to the boosted wavefunctions via
\begin{equation}
\langle J_\alpha({\bf r}) \rangle' = \langle \hat{J}^{\rm loc}_\alpha({\bf r}) \rangle' + \langle \hat{J}^{\rm nl}_\alpha({\bf r}) \rangle'.
\end{equation}
It is easy to show that 
\begin{align}
\langle \hat{J}^{\rm loc}_\alpha({\bf r}) \rangle' =& \langle \hat{J}^{\rm loc}_\alpha({\bf r-v }t) \rangle - n({\bf r-v}t)v_\alpha, \\
\langle \hat{J}^{\rm nl}_\alpha({\bf r}) \rangle' =& \langle \hat{J}^{\rm nl}_\alpha({\bf r-v}t) \rangle.
\end{align}
(The second expectation value is taken with the primed nonlocal operator and the primed wave functions; the additional phases in both cancel out, and we recover the nonlocal current in the unprimed system.) At the end of the day, we obtain 
\begin{equation}
{{\bf j}^{\rm el}}'({\bf r}) = {\bf j}^{\rm el}({\bf r-v}t) - n({\bf r-v}t){\bf v},
\end{equation}
which is the correct transformation law for the electronic contribution to the current density, ${\bf j}^{\rm el}$. By incorporating the 
ionic point-charge contribution we arrive at the transformation law for the total current density stated in the main text.

\subsection{ {Basis sets}}

 {Our arguments regarding the velocity dependence of the pseudopotentials are completely general and applicable to 
any basis set. The numerical implementation presented in the main text is based on norm-conserving pseudopotentials and 
plane waves. In this Section we briefly discuss the two most obvious alternatives: 
projector-augmented wave (PAW) and localized-orbital bases.}

\subsubsection{ {PAW}}

 {The PAW method~\cite{bloechl:1994} rests on a linear transformation between the pseudo (PS) 
and all-electron (AE) wave functions, $|\psi \rangle = \mathcal{T} |\tilde{\psi} \rangle$.
The operator $\mathcal{T}$ is defined as
\begin{equation}
\label{paw}
\mathcal{T} = {\bf I} + \sum_{\bf R,\nu} \left[|\phi_{\bf R \nu}\rangle -  |\tilde{\phi}_{\bf R \nu}\rangle \right] \langle \tilde{p}_{\bf R \nu}|
\end{equation}
which depends on a set of  are AE partial waves, $|\phi_{\bf R \nu}\rangle$,
together with the corresponding PS partial waves, $|\tilde{\phi}_{\bf R \nu}\rangle$ and projectors
$\langle \tilde{p}_{\bf R \nu}|$.
(${\bf R}$ refers to the atomic location, while $\nu$ is an index running over the total number of projectors 
at a given atomic site.)
Based on this mapping, the PS representation of an arbitrary semi-local operator $O$ is written as
\begin{equation}
\tilde{O} = O + \sum_{\bf R,\mu \nu} |\tilde{p}_{\bf R \mu}\rangle \left[ 
  \langle {\phi}_{\bf R \mu}|   O  |\phi_{\bf R \nu}\rangle -     \langle \tilde{\phi}_{\bf R \mu}| O  |\tilde{\phi}_{\bf R \nu}\rangle \right] \langle \tilde{p}_{\bf R \nu}|.
\end{equation}  
}

 {Our traveling pseudopotential method can be easily implemented, in the PAW context, by 
applying the same velocity-dependent phase to all individual objects in Eq.~\eqref{paw}.
This means redefining the transformation between PS and AE wavefunctions as
\begin{equation}
\label{paw_vel}
\mathcal{T} = {\bf I} + \sum_{\bf R,\nu} e^{i {\bf \dot{R} \cdot r}} \left[|\phi_{\bf R \nu}\rangle -  |\tilde{\phi}_{\bf R \nu}\rangle \right] \langle \tilde{p}_{\bf R \nu}|e^{-i {\bf \dot{R} \cdot r}}
\end{equation}
Then, the PS representation of a generic semi-local operator becomes
\begin{equation}
\begin{split}
\label{paw_velo}
\tilde{O} =& O + \sum_{\bf R,\mu \nu} e^{i {\bf \dot{R} \cdot r}}  |\tilde{p}_{\bf R \mu}\rangle  \\
& \times \Big[  \langle {\phi}_{\bf R \mu}| e^{i {\bf \dot{R} \cdot r}}   O e^{-i {\bf \dot{R} \cdot r}} |\phi_{\bf R \nu}\rangle \\
& -     \langle \tilde{\phi}_{\bf R \mu}| e^{i {\bf \dot{R} \cdot r}} Oe^{-i {\bf \dot{R} \cdot r}}  |\tilde{\phi}_{\bf R \nu}\rangle \Big] \\ 
&\times \langle \tilde{p}_{\bf R \nu}| e^{-i {\bf \dot{R} \cdot r}}.
\end{split}
\end{equation}  
Note the close similarity between the latter formulas and the gauge-including
projector-augmented wave (GIPAW) method of Pickard and Mauri~\cite{Pickard-01}. In particular, 
Eq.~\eqref{paw_vel} and Eq.~\eqref{paw_velo} can be directly obtained
from Eq.(16) and (17) of Ref.~\cite{Pickard-01} by replacing ${\bf R} \times {\bf B} / 2c$
(value of the electromagnetic vector potential at the atomic site) with $\dot{\bf R}$
(atomic velocity).
It is straightforward to show that Eq.~\eqref{paw_velo} transforms correctly under Galilean boosts.
The proof follows the same steps as for demonstrating the invariance of the GIPAW Hamiltonian
with respect to a shift of the electromagnetic gauge origin.}

\subsubsection{ {Localized orbitals}}

 {The most popular choices for localized orbital bases nowadays revolve around two
families of atom-centered functions: numerical atomic orbital (NAO) bases, 
as implemented in the SIESTA code, and Gaussian functions (GAUSSIAN, CP2K).
They both share one common feature: all basis functions, just like the atomic pseudopotentials,
are attached to a given atomic site, and displace rigidly with the atom they belong to during
the dynamical evolution of the system.
This means that, in principle, \emph{both} the basis functions and the nonlocal pseudopotential 
operators should be modified to account for the nuclear velocity.
%
%As a matter of fact, 
{As for the basis functions, the appropriateness of doing so
was already recognized long ago, e.g. in the simulation of high-enegy atomic collisions 
with traveling-orbital bases~\cite{Bates1953,Crothers1979,Kimura1985}, or
in Nafie's~\cite{Nafie-92} velocity-gauge approach to %In his approach to 
vibrational circular dichroism (VCD). 
In the context of nonlocal pseudopotentials, the use of velocity-gauge factors appears only
sparsely documented:
The only instance we could find in the literature is Ref.~\cite{Ditler-22},
describing a recent numerical implementation of Nafie's 
nuclear velocity perturbation theory.
%
%In a recent code implementation of the latter,~\cite{Ditler-22} nonlocal projectors 
%were transformed as well, although 
However, the general relevance of transforming the nonlocal projectors (i.e., beyond
the specifics of the Gaussian basis set used therein) was not understood.
%motivation for doing so appears to 
%molecules he introduced velocity-gauge factors in the atomic orbitals that %and projectors that
%closely resemble those discussed here.
%
% with analogous velocity-gauge factors were also adopted 
%
%
%as a matter of fact, 
%The only instance we could find in the literature is Eq.~(32) of Ref.~\cite{Ditler-22},
%
%
%This appears to be intended as a technical detail of the implementation
%but no formal justification or references are provided in support. 
%with no claims of generality beyond the specifics of the 
%Gaussian basis set used therein.
%
In the following, we demonstrate the key importance of both 
prescriptions in guaranteeing Galilean covariance of the time-dependent Schr\"odinger equation
within a localized-basis representation.}

%No formal justification or reference is provided in support .)
%we understand that 
%the general relevance of this prescription (i.e., beyond the  
%A numerical implementation in CP2K of the resulting ``nuclear velocity perturbation theory''
%is described in a recent work by Ditler et al.~\cite{Ditler-22}.}

% {To the best of our knowledge, the importance of the velocity-gauge factors was 
%only emphasized in the context of \emph{localized basis sets}, 
%e.g., the aforementioned perturbative approach to VCD,
%or 
%
%That is, the general relevance of this prescription to nonlocal pseudopotentials, 
%and its key role in guaranteeing Galilean covariance of the time-dependent Schr\"odinger equation, 
%was never pointed out: we shall demonstrate the latter point in the following.
%}

 {The time-dependent Schr\"odinger equation for an evolving nonorthogonal 
basis set can be written as~\cite{Todorov2001}
\begin{equation}
\label{tdse}
\sum_\nu (H_{\mu\nu} - i D_{\mu\nu} ) C_\nu = i \sum_\nu S_{\mu \nu} \dot{C}_\nu,
\end{equation}
where $|\psi \rangle = \sum_\nu |e_\nu\rangle C_\nu$ is an electronic wave-function
(we omit the band index $n$ to simplify the notation) and the Greek indices run over 
the NAO or Gaussian basis. $H_{\mu \nu} = \langle e_\mu | H |e_\nu\rangle$ is
the Hamiltonian represented on such basis, $D_{\mu\nu} =  \langle e_\mu |  \partial_t e_\nu\rangle$,
and $S_{\mu\nu} =  \langle e_\mu | e_\nu\rangle$.}

 {We now use the following \emph{Ansatz} for the basis functions,
\begin{equation}
\label{trav_orb}
\langle {\bf r}| e_\nu\rangle = e^{ i \dot{\bf R}_\nu \cdot ({\bf r} - {\bf R}_\nu)} \phi_\nu ({\bf r} - {\bf R}_\nu),
%\langle {\bf r}| e_\nu\rangle = e^{ i \dot{\bf R}_\nu \cdot {\bf r}} \phi_\nu ({\bf r} - {\bf R}_\nu),
\end{equation}
where ${\bf R}_\nu$ is the instantaneous location of the atom that the element $\nu$ of
the basis belongs to. We then proceed to demonstrating that Eq.~\eqref{trav_orb}, in
conjunction with our traveling pseudopotential prescription [Eq.(5) of the main text]
is sufficient to guarantee the exact Galilean covariance of Eq.~\eqref{tdse}.}

 {After some algebra, we find the following transformation laws for the momentum,
kinetic energy, external potential and overlap operators,
\begin{subequations}
\begin{align}
 {\bf p}'_{\mu \nu}  =&  e^{ i {\bf v} \cdot ({\bf R}_\mu - {\bf R}_\nu)} ( {\bf p}_{\mu\nu} + m{\bf v} S_{\mu\nu}), \\
T'_{\mu \nu} =&e^{ i {\bf v} \cdot ({\bf R}_\mu - {\bf R}_\nu)}  \left( T_{\mu \nu}  + {\bf v} \cdot {\bf p}_{\mu \nu} + \frac{m}{2} v^2 S_{\mu\nu} \right), \\ 
V^{\rm ext'}_{\mu \nu} =& e^{ i {\bf v} \cdot ({\bf R}_\mu - {\bf R}_\nu)}  V^{\rm ext}_{\mu\nu}, \label{vext} \\
S'_{\mu \nu} =& e^{ i {\bf v} \cdot ({\bf R}_\mu - {\bf R}_\nu)}  S_{\mu \nu}.
\end{align}
\end{subequations}
(For a given operator $O$ we define $O_{\mu \nu} =  \langle e_\mu | O | e_\nu\rangle$.) 
Note that the transformation law for the external potential, Eq.~\eqref{vext} \emph{requires} using
traveling pseudopotentials, or this relation (and hence the remainder of the derivation) would break down.}

 {Based on the above relations, and neglecting SCF fields for the moment,
we obtain the following relation between the original and 
boosted Hamiltonians,
\begin{equation}
H'_{\mu \nu} = e^{ i {\bf v} \cdot ({\bf R}_\mu - {\bf R}_\nu)}  \left( H_{\mu \nu}  + {\bf v} \cdot {\bf p}_{\mu \nu} + \frac{m}{2} v^2 S_{\mu\nu} \right).
\end{equation}
Regarding the connection $iD_{\mu \nu}$, the Galileo transformation yields the following result,
\begin{equation}
\begin{split}
& -i D'_{\mu\nu} =  e^{ i {\bf v} \cdot ({\bf R}_\mu - {\bf R}_\nu)}  \\
 &\quad \times \Big( -i D_{\mu\nu} - m \dot{\bf R}_\nu \cdot {\bf v} S_{\mu \nu} - mv^2 S_{\mu \nu} - {\bf v} \cdot {\bf p}_{\mu \nu} \Big).
\end{split}
\end{equation}
Thus, we have
\begin{equation}
\begin{split}
& H'_{\mu \nu} -i D'_{\mu\nu} =  e^{ i {\bf v} \cdot ({\bf R}_\mu - {\bf R}_\nu)}  \\
 &\quad \times \Big( H_{\mu \nu}-i D_{\mu\nu} - m \dot{\bf R}_\nu \cdot {\bf v} S_{\mu \nu} - \frac{mv^2}{2} S_{\mu \nu} \Big).
\end{split}
\end{equation}
}

 {Based on the above results, we postulate the following transformation law for the wave-function coefficients,
\begin{equation}
\label{cprime}
C'_\nu = e^{ i {\bf v} \cdot {\bf R}_\nu + i mv^2 t/2} C_\nu
\end{equation} 
The time derivative reads as
\begin{equation}
\begin{split}
\dot{C}'_\nu %=& e^{ i {\bf v} \cdot {\bf R}_\nu + i mv^2/2 t} \dot{C}_\nu + i{\bf v} \cdot \dot{\bf R}_\nu e^{ i {\bf v} \cdot {\bf R}_\nu + i mv^2/2 t} C_\nu
% + i mv^2/2 e^{ i {\bf v} \cdot {\bf R}_\nu + i mv^2 t/2} C_\nu \\
 =& e^{ i {\bf v} \cdot {\bf R}_\nu + i mv^2t/2} \left( \dot{C}_\nu + i{\bf v} \cdot \dot{\bf R}_\nu C_\nu + i \frac{mv^2}{2} C_\nu \right).
 \end{split}
\end{equation} 
This means that the rhs of the boosted TDSE is
\begin{equation}
\begin{split}
i S'_{\mu \nu} \dot{C}'_\nu =& e^{ i {\bf v} \cdot {\bf R}_\mu + i mv^2 t /2} \\
& \times \Big( i S_{\mu \nu}  \dot{C}_\nu - S_{\mu \nu} {\bf v} \cdot \dot{\bf R}_\nu C_\nu -
 \frac{mv^2}{2} S_{\mu \nu} C_\nu \Big).
\end{split}
\end{equation}
The additional pieces cancel out with the extra terms in $H'_{\mu \nu} - i D'_{\mu \nu}$, and we recover the TDSE in the unboosted state.
In other words, if $C_\nu$ is a solution of Eq.~\eqref{tdse}, then $C'_\nu$ as defined in Eq.~\eqref{cprime} is a solution
of the boosted TDSE, $\sum_\nu (H'_{\mu\nu} - i D'_{\mu\nu}) C'_\nu = i \sum_\nu S'_{\mu \nu} \dot{C}'_\nu$.}

 {It is straightforward at this point to prove that the electron density (and hence the SCF potential) complies with the rule
\begin{equation}
n'({\bf r},t) = n({\bf r}- {\bf v}t, t),
\end{equation}
which proves the Galilean covariance of the interacting Hamiltonian.
It is equally straightforward to prove that Eq.~\eqref{banden} holds for the present case of a localized basis set.
Note that we never invoked basis set completeness in the derivation, which means that Galilean covariance is
guaranteed even for a minimal basis of tight-binding-like atomic orbitals.}

%\begin{equation}
%H'_{\mu \nu} -i D'_{\mu\nu} = H_{\mu \nu} -i D_{\mu\nu} - m \dot{\bf R}_\nu \cdot {\bf v} S_{\mu \nu} - \frac{mv^2}{2} S_{\mu \nu}.
%\end{equation}

 {To complete the above derivations, it is interesting to go beyond the case of Galileo transformation, and
consider a generalized boost with a velocity ${\bf v}$ that is not constant in time.
In such a situation, one can show that $C'_\nu$ as defined in Eq.~\eqref{cprime} solves the TDSE in the boosted system if $C_\nu$ 
is a solution of the following TDSE,
\begin{equation}
\label{tdse2}
\sum_\nu (H_{\mu \nu} + \dot{\bf v} \cdot {\bf r}_{\mu \nu} - i D_{\mu\nu} ) C_\nu = i \sum_\nu S_{\mu \nu} C_\nu.
\end{equation}
(Here ${\bf r}_{\mu \nu} = \langle e_\mu | {\bf r} | e_\nu\rangle$ are the matrix elements of the position operator
on the localized basis set.)
This means that a time-dependent acceleration has the exact same physical consequences 
as a time-dependent electric field, consistent with the arguments of the main text.}

\section{Linear response}

\subsection{Force constants}

The electronic contribution to the dynamical force constants can be written as (we follow Gonze's convention~\cite{gonze/lee} and indicate the phonon perturbation 
with the label $\tau_{\kappa\alpha}$)
\begin{equation}
C^{\rm el}_{\kappa \alpha,\kappa'\beta} (\w) = {\rm Tr} \, \left[ \hat{H}^{\tau_{\kappa \alpha}}(-\w)  \hat{\rho}^{\tau_{\kappa' \beta}}(\w) + \hat{H}^{\tau_{\kappa \alpha}\tau_{\kappa' \beta}}  \hat{\rho} \right],
\end{equation}
Here $\hat{H}^{\tau_{\kappa \alpha}}$ and $\hat{H}^{\tau_{\kappa \alpha}\tau_{\kappa' \beta}}$ are, respectively, 
the first and second derivatives of the external potential with respect to the atomic coordinates, while $\hat{\rho}$
and $ \hat{\rho}^{\tau_{\kappa' \beta}}(\w)$ are the ground-state and first-order density operator. 
The latter is related to the first-order Hamiltonian via the Liouville equation at linear order, which for an arbitrary 
perturbation parameter $\lambda$ reads as
\begin{equation}
\w \hat{\rho}^{\lambda} = [\hat{H},\hat{\rho}^{\lambda}] + [\hat{H}^{\lambda},\hat{\rho}].
\end{equation}

By considering the velocity dependence of the pseudopotentials, the first-order Hamiltonian acquires the following form,
\begin{equation}
\label{dHdu}
%\frac{\partial \hat{H}}{\partial u_{\kappa\alpha}} 
\hat{H}^{\tau_{\kappa \alpha}}(\w) = \frac{\partial \hat{V}^{\rm ps}}{\partial R_{\kappa\alpha}} % \hat{H}^{\tau_{\kappa\alpha}} 
- i\w \frac{\partial \hat{V}^{\rm ps}}{\partial \dot{R}_{\kappa \alpha}}.
\end{equation}
where $\hat{V}^{\rm ps}$ is the total pseudopotential contribution to the Hamiltonian, including the local part. The first term in Eq.~\eqref{dHdu} is the standard phonon Hamiltonian~\cite{gonze/lee,Baroni-01}, independent of frequency; the second term originates from the velocity-dependent nature of the nonlocal pseudopotential operator, and introduces an explicit frequency dependence in the external perturbation.
More explicitly, we have
\begin{subequations}
\begin{align}
\frac{\partial  V^{\rm ps}}{\partial {\bf R}_\kappa} = &  i \left[ V^{\rm ps}_\kappa,{\bf p}  \right], \\
\frac{\partial  V^{\rm ps}}{\partial \dot{\bf R}_\kappa} =& -  i \left[ V^{\rm nl}_\kappa, {\bf r}  \right].
\end{align}
\end{subequations}
In reciprocal space, the commutator with ${\bf r}$ becomes a derivative with respect to crystal momentum, and
the first-order Hamiltonian becomes 
\begin{equation}
 \hat{H}^{ \tau_{\kappa\alpha}}_{\bf k}(\w) = \frac{\partial \hat{V}^{\rm ps} ({\bf k})}{\partial R_{\kappa\alpha}} + 
i\w \frac{\partial \hat{V}_{\kappa}^{\rm nl}({\bf k})}{\partial k_\alpha}.
\end{equation}

For the second-order Hamiltonian, we have
\begin{equation}
\label{h2}
\begin{split}
%\frac{\partial^2 \hat{H}}{\partial u_{\kappa\alpha} \partial u_{\kappa' \beta}} 
\hat{H}^{\tau_{\kappa \alpha}\tau_{\kappa' \beta}}  =& \delta_{\kappa \kappa'} \Bigg(
  \frac{\partial^2 \hat{H}}{\partial R_{\kappa\alpha} \partial R_{\kappa \beta}} \\
 &  - i\w  \frac{\partial^2 \hat{V}^{\rm nl}_\kappa }{ \partial k_{\alpha} \partial R_{\kappa \beta}} 
  + i\w  \frac{\partial^2 \hat{V}^{\rm nl}_\kappa }{\partial R_{\kappa \alpha} \partial k_{\beta}}  \\
&   + \w^2 \frac{\partial^2 \hat{V}^{\rm nl}_\kappa }{\partial k_\alpha \partial k_\beta} \Bigg).
\end{split}   
\end{equation}
We have a total of four terms: the first line coincides with the usual static term of the standard implementations; the second line 
contains terms that occur at first order in $\w$; finally, the third line entails a contribution that is proportional to $\w^2$. 
The terms that are linear in the frequency are relevant only in crystals that break TR invariance;
in these situations, we expect these to contribute to the molecular Berry curvature. 
In the materials set that we tested in this work, the correction to 
the second-order phonon Hamiltonian due to traveling pseudopotentials reduces to
the bottom line in Eq.~\eqref{h2}.

\subsection{Electric fields} 

Before moving to the Born charges, we briefly recap the linear response to a
uniform electric field within density-functional perturbation theory.
Consider an external electric field that is uniform in space and oscillating in time as
\begin{equation}
\mathcal{E}_\alpha(t) = \mathcal{E}_\alpha e^{-i\w t}.
\end{equation}
Within time-dependent density-functional perturbation theory, 
we can write the matrix elements of the first-order density operator in the following form,
\begin{equation}
\label{paSCF}
\begin{split}
 \langle m| \hat{\rho}^\Ea(\w)|n\rangle =  & i \frac{\bar{f}_{nm} \langle m | H^\alpha |n\rangle}{\e_n-\e_m + \omega}  \\
  &    + \frac{(f_n-f_m) \langle m | V^\E_\alpha(\w) |n\rangle }{\e_n-\e_m + \omega}.
\end{split}
\end{equation}      
Here $|n\rangle$ stands for an unperturbed ground-state wave function with eigenvalue $\e_n$ and occupancy $f_n$; 
$H^\alpha = i[H,r_\alpha]$ is the velocity operator;
$V^\E_\alpha(\w)$ describes the self-consistent Hartree and exchange and correlation fields.
Finally, we have introduced the following function of the band energies
\begin{equation}
\bar{f}_{mn} = \frac{f_n-f_m}{\e_n-\e_m} \, \, (m\neq n), \qquad  \bar{f}_{nn} = f'(\e_n).
\end{equation}
(The fraction is meant to be replaced smoothly with the first derivative of the occupation 
function whenever the energy difference at the denominator becomes small.)

\subsection{Born charges}

 We start by addressing the Born dynamical charges, which we write as the atomic forces induced at linear order by an applied electric field $\E_\beta$,
 \begin{equation}
 %\frac{\partial f_{\kappa\alpha}}{\partial A_\beta} 
 \label{zstar}
 Z_{\kappa\alpha,\beta}= \frac{\partial f_{\kappa\alpha}}{\partial \E_\beta} = 
 Z_\kappa \delta_{\alpha \beta} - \lim_{\w \rightarrow 0} {\rm Tr} \Big( \hat{H}^{\tau_{\kappa \alpha}}(-\w) \hat{\rho}^{\E_\beta}(\w) \Big).
 %+
 %\hat{H}^{u_{\kappa \alpha} k_\beta}(-\w) \hat{\rho}  \Big).
 \end{equation}
The first term depends on the bare pseudopotential charge in the usual manner~\cite{Baroni-01,gonze/lee}, while the electronic contribution is governed by 
the first-order density operator with respect to $\E_\beta$. 

In insulators, the $\w\rightarrow 0$ limit of Eq.~\eqref{zstar} goes smoothly, and yields the established formula for the Born effective charges. In other words, the second term in Eq.~\eqref{dHdu} does not contribute in the static limit, and the existing DFPT implementations for insulators are unaffected by our prescriptions.
[Note that the velocity dependence of the nonlocal operators \emph{does} contribute whenever the Born charges are calculated dynamically, and is essential to guarantee that Eq.({15b}) of the main text is fulfilled exactly at any frequency.]

In metals, on the other hand, the intraband part of $\hat{\rho}^{\E_\beta}(\w)$ diverges in the low-frequency limit,
\begin{equation}
\hat{\rho}^{\E_\beta, {\rm ra}}(\w) = \frac{i}{\w} \sum_n |n\rangle \frac{\partial f_n}{\partial k_\beta} \langle n|.
\end{equation}
This yields an additional contribution to the dynamical charges that tends to a constant in the static limit, 
\begin{equation}
\Delta Z_{\kappa\alpha,\beta}= 
\sum_n \frac{\partial f_n}{\partial k_\beta} \langle n|  \frac{\partial \hat{V}^{\rm nl}}{\partial \dot{R}_{\kappa \alpha}} |n\rangle.
\end{equation}
Now, recall the sum rule established in Ref.~\cite{DreyerPRL22} for the uncorrected dynamical charges (i.e., those obtained while neglecting the velocity dependence of the nonlocal pseudopotential operator),
\begin{equation}
\sum_\kappa \tilde{Z}_{\kappa\alpha,\beta}= - 
\sum_n \frac{\partial f_n}{\partial k_\beta} \langle n|  \hat{p}_\alpha |n\rangle.
\end{equation}
By recalling the following relation,
\begin{equation}
\label{vsum}
\sum_\kappa \frac{\partial \hat{V}^{\rm nl}}{\partial \dot{R}_{\kappa \alpha}} = -\frac{\partial \hat{V}^{\rm nl}}{\partial k_{\alpha}},
\end{equation}
we arrive immediately at the following sum rule for the corrected dynamical charges, ${Z}_{\kappa\alpha,\beta} = \tilde{Z}_{\kappa\alpha,\beta} 
+ \Delta Z_{\kappa\alpha,\beta}$,
\begin{equation}
\sum_\kappa {Z}_{\kappa\alpha,\beta}= - 
\sum_n \frac{\partial f_n}{\partial k_\beta} \langle n| \hat{H}^{k_\alpha} |n\rangle = -\sum_n f'_n v_n^{(\alpha)}  v_n^{(\beta)}, 
\end{equation}
where $v_n^{(\alpha)}= \langle n| \hat{H}^{k_\alpha} |n\rangle$ is the band velocity.
The quantity on the rhs now coincides with $\Omega D_{\alpha\beta} / \pi$, where $\Omega$ is the cell volume and 
$D_{\alpha\beta}$ is the symmetric Drude weight tensor as derived from the low-frequency limit of the optical conductivity. 
This result, Eq.~(18) of the main text,
fixes an undesirable inconsistency of the existing 
linear-response theory, which remained unresolved to date. 

\subsection{Translations}

The above result can be better understood by considering the linear response of the  
crystal to a rigid translation of the whole lattice, ${\bf u}$.
The corresponding first-order Hamiltonian is given by the sublattice sum of Eq.~\eqref{dHdu},
\begin{equation}
\hat{H}^{u_{\beta}}(\w) = \sum_\kappa \hat{H}^{\tau_{\kappa \alpha}}(\w) =  i [\hat{H},\hat{p}_\beta]
+ i\w \frac{\partial \hat{V}^{\rm nl}}{\partial k_{\beta}},
\end{equation}
where the first term stems from translational invariance of the ground-state Hamiltonian~\cite{SchiaffinoPRB19,DreyerPRL22} 
and the second term is a consequence of Eq.~\eqref{vsum}.
The derivative of the density operator with respect to $u_\beta$ then reduces to
\begin{equation}
\hat{\rho}^{u_\beta}(\w) = \hat{\rho}^{u_\beta} + i \w \hat{\rho}^{A_\beta}(\w),
\end{equation}
where $\hat{\rho}^{\bf u}= i[\hat{\rho},\hat{\bf p}]$ describes the response to a static translation, 
and the dynamical part \emph{coincides} with the response to a vector potential field that is modulated in time.
Thus, our formalism restores the exact equivalence between electric fields and 
mechanical boosts that was derived in Ref.~\cite{StengelPRB18} for a local Hamiltonian. 
This equivalence is, of course, violated if the
velocity-dependent character of the nonlocal pseudopotential operators is neglected in the dynamics,
which proves the necessity of Eq.~(5) of the main text on general physical grounds.

Now we use the above result to calculate the second derivative of the action
with respect to $\tau_{\kappa\alpha}$ and $u_\beta$, corresponding physically to
minus the force induced on the sublattice $\kappa$ along $\alpha$ by the
time-dependent boost $u_\beta$.
By construction, this also corresponds to the sum with respect to the sublattice 
index $\kappa'$ of the nonadiabatic force constants $C_{\kappa \alpha,\kappa' \beta}(\w)$.
We have
\begin{equation}
E^{\tau_{\kappa \alpha} u_\beta} = {\rm Tr}  \Big( \hat{H}^{u_{\kappa \alpha}} \hat{\rho}^{u_\beta}  +  \hat{H}^{u_{\kappa \alpha}u_\beta}
\hat{\rho}  \Big).
\end{equation}
After observing that 
\begin{equation}
\hat{\rho}^{A_\beta}(\w) = \hat{\rho}^{k_\beta} + i\w \hat{\rho}^{\E_\beta}(\w),
\end{equation}
we obtain
\begin{equation}
\hat{\rho}^{u_\beta}(\w) = \hat{\rho}^{u_\beta} + i \w \hat{\rho}^{k_\beta} - \w^2 \hat{\rho}^{\E_\beta}(\w).
\end{equation}
The contributions of the first two terms cancel out exactly with those of the second-order Hamiltonian, 
and we are left with the contribution of the electric field density operator.
We find
\begin{equation}
\begin{split}
\sum_{\kappa'} C_{\kappa \alpha,\kappa' \beta}(\w) =& -\w^2 {\rm Tr } \Big( \hat{H}^{u_{\kappa \alpha}}(-\w) \hat{\rho}^{\E_\beta}(\w) \Big) \\
 =& \w^2 Z^{\rm el}_{\kappa\alpha,\beta}(\w). % \Big( Z^*_{\kappa\alpha,\beta}(\w) - Z_\kappa \delta_{\alpha \beta} \Big),
\end{split}
%(Z_{\kappa \alpha,\beta} - N
\end{equation}
i.e., at second order in the frequency we obtain the electronic contribution to the Born dynamical charges, 
consistent with Eq.~(15a) of the main text.
By summing the force constants over \emph{both} sublattices we find
\begin{equation}
\label{sum2}
\begin{split}
\sum_{\kappa \kappa'} C_{\kappa \alpha,\kappa' \beta}(\w)  =& \w^2 \Big(\frac{\Omega D_{\alpha\beta}}{\pi} - N \delta_{\alpha \beta} \Big) + O(\w^3),
\end{split}
%(Z_{\kappa \alpha,\beta} - N
\end{equation}
This result coincides with Eq.~(19) of the main text.

\subsection{Rotations}

The treatment of rotations follows the same guidelines as that of translations presented earlier in this document.
The derivative of the Hamiltonian with respect to an infinitesimal rotation of an isolated object is
\begin{equation}
\begin{split}
\hat{H}^{\bm{\theta}}(\w) =& \sum_\kappa {\bf R}_\kappa \times \hat{H}^{\bm{\tau}_{\kappa}}(\w) \\
 =&  i[\hat{H}, {\bf r} \times {\bf p}] + i\w \sum_\kappa {\bf R}_\kappa \times  i \left[ V^{\rm nl}_\kappa, {\bf r}  \right].
\end{split}
\end{equation}
The first-order density matrix can be written as
\begin{equation}
\hat{\rho}^{\bm{\theta}}(\w) = i [\hat{\rho},{\bf r} \times {\bf p}] + i\w \hat{\rho}^{\bf L}(\w).
\end{equation}
The frequency-dependent part $\hat{\rho}^{\bf L}(\w)$ consists in the response to the following perturbing 
operator,
\begin{equation}
\hat{\bf L}  = {\bf r} \times {\bf p} + \sum_{\kappa} {\bf R}_{\kappa} \times i \left[ V^{\rm nl}_\kappa, {\bf r}  \right].
\end{equation}
This is, of course, the orbital angular momentum operator introduced in Eq.~(10) of the main text,
which is related to the first-order GIPAW~\cite{Pickard-03} Hamiltonian by a factor of $1/2c$.
The present derivation proves the equivalence between a rotation with uniform angular velocity
and an external magnetic field up to first order. In other words, our traveling pseudopotentials
restore the exact validity of the Larmor theorem even when the electron-nuclei interaction
is described by nonlocal operators. 

\section{Adiabatic dynamics}

\subsection{Berry connection}

 {We provide here a derivation of Eq.~(11) of the main text. We follow the general guidelines of Section 3.3 of Ref.~\cite{David-book}, 
which we generalize to the case of a Hamiltonian that explicitly depends on velocity.
As in Ref.~\cite{David-book}, we assume for simplicity a single isolated band $|\psi \rangle$ that evolves adiabatically 
along a structural path in configuration space, which we indicate with the parameter $\lambda$. We use the ansatz
\begin{equation}
\label{ansatz}
|\psi(t)\rangle = c(t) e^{-i\gamma(t)} |n(t)\rangle.
\end{equation}
Here $|n(t)\rangle=|n(\lambda(t))\rangle$ is the instantaneous eigenstate of the time-independent Hamiltonian,
with eigenvalue $E_n(t)$; 
\begin{equation}
\gamma(t) = \frac{1}{\hbar} \int_0^t E_n(t') dt'
\end{equation}
is the dynamical phase factor, and $c(t)$ is related to the Berry connection.}

 {To determine $c(t)$, we plug Eq.~\eqref{ansatz} into the time-dependent Schr\"odinger equation (TDSE), 
\begin{equation}
[i\hbar \partial_t - H(t)] |\psi(t)\rangle = 0.
\end{equation}
It is important to observe, at this point, that $H(t)$ does \emph{not} coincide with the static Hamiltonian at time $t$, $H^{\rm st}(\lambda(t))$,
since it depends explicitly on the instantaneous rate of change of $\lambda$.
%which we have used to define the Born-Oppenheimer state $|n(t)\rangle$. 
%
At lowest order in the
velocity, we can write
\begin{equation}
H(t) \simeq H^{\rm st}(\lambda(t)) + \dot{\lambda} \frac{\partial H(\lambda(t))}{\partial \dot{\lambda}} \Big|_{\dot{\lambda}=0},
\end{equation}
where the first term on the rhs defines the instantaneous eigenstate via
\begin{equation}
H^{\rm st}(\lambda)|n(\lambda)\rangle =  E_n(\lambda)|n(\lambda)\rangle,
\end{equation}
and the reminder embodies the velocity dependence.
Thus, the TDSE reads as
\begin{equation}
i \dot{c}(t) |n(t)\rangle + i c(t) \partial_t   |n(t)\rangle -  \dot{\lambda} \frac{\partial H}{\partial \dot{\lambda}} |n(t)\rangle = 0.
\end{equation}
By projecting on $\langle n(t)|$, we obtain
\begin{equation}
\dot{c}(t) = i c(t) A_n(t),
\end{equation}
with
\begin{equation}
\begin{split}
A_n(t) = &\langle n(t) | i \partial_t n(t) \rangle - \dot{\lambda} \langle n(t) |  \partial_{\dot{\lambda}} H |n(t) \rangle, \\
   =& \dot{\lambda} \left( \langle n(\lambda) | i \partial_\lambda n(\lambda) \rangle  -  \langle n(\lambda) |  \partial_{\dot{\lambda}} H |n(\lambda) \rangle \right).
\end{split}
\end{equation}
Given that $A^{\rm nl}_\lambda = -  \langle n(\lambda) |  \partial_{\dot{\lambda}} H |n(\lambda) \rangle$, this result coincides with Eq.~(11) of the main text.}

\subsection{Berry curvatures}

 {Our goal in this section is to provide an alternative derivation of one of our formal results {(see Sec.~III.C)}, that the adiabatic
charge transport in insulators is unaffected, at leading order, by the velocity dependence of the nonlocal pseudopotentials. 
To that end, we consider the Berry curvature with respect to crystal momentum and atomic displacement, which defines~\cite{Resta-25}
the electronic contribution to the Born effective charges in the zero-frequency limit.
For the Berry connection in configuration space we use Eq.(11) of the main text, while in crystal momentum space we have as usual
\begin{equation}
A_{\beta} =  i \sum_n \Big\langle \psi_n \Big| \frac{\partial \psi_n}{\partial k_{\beta}} \Big\rangle.
\end{equation}
This means that the Berry curvature is
\begin{equation}
\label{omega}
\begin{split}
\Omega_{\kappa\alpha,\beta} =& \frac{\partial A_{\kappa\alpha}}{\partial k_{\beta}} - \frac{\partial A_{\beta}}{\partial R_{\kappa\alpha}} \\
  =& i  \sum_n \Bigg( \Big\langle  \frac{\partial \psi_n}{\partial k_{ \beta}} \Big| \frac{\partial \psi_n}{\partial R_{\kappa \alpha}} \Big\rangle -
    \Big\langle  \frac{\partial \psi_n}{\partial R_{\kappa \alpha}} \Big| \frac{\partial \psi_n}{\partial k_{\beta}} \Big\rangle \Bigg) 
    + \frac{\partial A^{\rm nl}_{\kappa\alpha}}{\partial k_{\beta}}.
\end{split}
\end{equation}
The last term embodies the traveling pseudopotential correction; however, this term is manifestly written as a total derivative in momentum space. In insulators, its Brillouin
zone integral vanishes identically, leaving the standard expression unaltered. This conclusion is nicely consistent with our linear-response results {in Sec. III.C of this document}}.

\begin{table}
\setlength{\tabcolsep}{6pt}
\begin{center}
\caption{\label{tab_geom} Relaxed geometries for the molecules studied in this work. 
Bond distances are reported in bohr.}

\begin{tabular}{r|c}\hline\hline
   \T\B & \multicolumn{1}{c}{Relaxed geometry} \\
   \hline
   \T\B  H$_2$      &  $d=1.444$  \\
   \T\B  N$_2$      &  $d=2.072$  \\
   \T\B  F$_2$      &  $d=2.628$  \\
   \T\B  P$_2$      &  $d=3.563$  \\
   \T\B  HF     &  $d=1.758$  \\
   \T\B  CH$_4$     &  $d=2.069$   \\
   \T\B  C$_6$H$_6$ &  $d_{\rm C-C}=2.622$, $d_{\rm C-H}=2.065$   \\
     \hline \hline
\end{tabular}
\end{center}
\end{table}

\section{Results}

\subsection{Computational details}

We perform our calculations using density functional theory (DFT) and density functional perturbation theory (DFPT) as implemented in ABINIT v10.1. Our approach builds upon the finite-frequency linear-response implementation used in Ref.~\cite{Royo-26}, which we extend by incorporating contributions arising from a velocity-dependent nonlocal operator, as discussed in this work.
The exchange-correlation functional is described using the Perdew-Wang parametrization of the Local Density Approximation (LDA) for both ground-state and linear-response calculations. We employ scalar relativistic norm-conserving pseudopotentials from the PseudoDojo~\cite{pseudodojo} library  {(standard level of accuracy)}.

\begin{table}[!b]
\setlength{\tabcolsep}{3pt}
\begin{center}
\caption{\label{eq20a_sto}Numerical test of the sum rule Eq.(15a) of the main text in bulk SrTiO$_3$. 
The first column corresponds to the usual definition of the Born charges in the static limit; second and third columns are calculated at  $\hbar \w=0.04$ Ha. Values in parentheses are obtained by neglecting  
the velocity dependence of $\hat{V}^{\rm nl}$. 
}
\begin{tabular}{c|rrr}\hline\hline
   \T\B & \multicolumn{1}{c}{$Z^{\rm el}_{\kappa\alpha,\alpha}(0)$} & \multicolumn{1}{c}{$Z^{\rm el}_{\kappa\alpha,\alpha}(\omega)$} &
   \multicolumn{1}{c}{$\frac{1}{\w^2}\sum\limits_{\kappa'} C_{\kappa \alpha,\kappa' \beta}(\w)$}  \\
   \hline
   \T\B Sr   & $-$7.4432   & $-$7.4521  ($-$7.4523)  &  $-$7.4521  ($-$7.4610) \\
   \T\B Ti   & $-$4.7165   & $-$4.3406  ($-$4.3441)  &  $-$4.3407  ($-$8.8157)  \\
   \T\B O1   & $-$8.0568   & $-$8.1901  ($-$8.1932)  &  $-$8.1894  ($-$8.1439)  \\
   \T\B O2   & $-$8.0568   & $-$8.1901  ($-$8.1932)  &  $-$8.1894  ($-$8.1439)  \\
   \T\B O3   & $-$11.7266  & $-$12.0972 ($-$12.1045) &  $-$12.0965 ($-$12.8290) \\
   \T\B sum  & $-$40.0000  & $-$40.2701 ($-$40.2872) &  $-$40.2682 ($-$45.3933) \\
  \hline \hline
\end{tabular}
\end{center}
\end{table}

For our calculation on bulk cubic crystals, we use a plane-wave cutoff of 80 Ha, and a $12\times12\times12$ Monkhorst-Pack ($\Gamma$-centered) $\bf k$-point mesh for GaP and diamond (SrTiO$_3$).
The cell parameters were previously optimized to the following values in bohr: $a_0$(SrTiO$_3$)= 7.2796, $a_0$(GaP)= 10.1813, $a_0$(C-diamond)= 6.6853. 
For isolated finite systems, we adopt large cubic cells to minimize interactions with periodic images. Specifically, we use a cell parameter of $a_0= 20$ bohr and a plane-wave cutoff of $E_{\rm cut}=100$ Ha in all cases, except for Ar ($a_0= 14$ bohr, $E_{\rm cut}=80$ Ha) and benzene ($a_0= 30$ bohr). %The Brillouin zone is sampled with a single $\Gamma$-point.
The data about the structure are summarized in Table~\ref{tab_geom}.

We perform dynamic DFPT calculations in the transparent regime. Specifically, we first carry out explicit calculations for atomic displacement and electric field perturbations at a discrete set of frequencies: $\hbar\omega$= 0.00, 0.01, 0.02, 0.03, 0.04, and 0.05 Ha.
Subsequently, we apply a polynomial interpolation to the computed linear-response quantities (IFCs, Born charges and dielectric susceptibilities), enabling the extraction of the expansion coefficients at the desired order in $\omega$.

\begin{table}
\setlength{\tabcolsep}{3pt}
\begin{center}
\caption{\label{eq20bc} Same as Table I of the main text, for isolated atoms and molecules.
Values in parentheses are obtained by neglecting  
the velocity dependence of $\hat{V}^{\rm nl}$. 
The first column serves as a reference, while the second and third columns are reported as relative and absolute deviations, respectively. Values in the third column should correspond to the number of valence electrons. 
}
\begin{tabular}{c|ccc}\hline\hline
   \T\B & \multicolumn{1}{c}{$-\omega^2\Omega\chi_{\alpha\alpha}(\omega)$} & \multicolumn{1}{c}{$\sum\limits_{\kappa} Z^*_{\kappa\alpha,\alpha}(\omega)$} &
   \multicolumn{1}{c}{$\frac{1}{\omega^2}\sum\limits_{\kappa,\kappa'} C_{\kappa\alpha,\kappa'\alpha}(\omega)$}  \\
   \hline
%   \T\B GaP       & $-$0.3507 & $<$0.01 (1.04)  & 18.000 (23.97) \\
%   \T\B SrTiO$_3$ & $-$0.2701 & $<$0.01 (6.33)  & 39.998 (45.12) \\
%   \T\B C         & $-$0.0452 &    0.01 (5.25)  & 8.000 (8.48) \\
   \T\B Ar        & $-$0.0197 &    0.02 (12.71) & 8.0000 (9.62)\\
   \T\B H$_2$     & $-$0.0121 &    0.25 (0.23)  & 2.0000 (2.00) \\
   \T\B N$_2$     & $-$0.0247 &    0.59 (2.78)  & 9.9994 (10.60) \\
   \T\B F$_2$     & $-$0.0199 & $<$0.01 (29.02) & 14.0005 (15.68) \\
   \T\B HF        & $-$0.0112 &    0.06 (15.32) & 8.0001 (8.56) \\
   \hline \hline
\end{tabular}
\end{center}
\end{table}

\begin{table}[!b]
\setlength{\tabcolsep}{6pt}
\begin{center}
\caption{\label{tab_magnetic} Same as Table II of the main text, for a different set of molecules. The rigid-core correction to the susceptibility has \emph{not} been included
in the reported values. It consists in an isotropic contribution, $\Delta \chi^{\rm mag}_{\alpha\beta} = \delta_{\alpha \beta} \chi^{\rm core}$, with $\chi^{\rm core}$ reported (units of $10^{-6}$ cm$^3$/mol) in the rightmost column. } %.
\begin{tabular}{r|rrrrr}\hline\hline
   \T\B & \multicolumn{1}{c}{$g$ factor} & \multicolumn{1}{c}{$\mathcal{\chi}^{\rm mag}_{xx}$} &
   \multicolumn{1}{c}{$\mathcal{\chi}^{\rm mag}_{yy}$} & \multicolumn{1}{c}{$\mathcal{\chi}^{\rm mag}_{zz}$} & $\chi^{\rm core}$ \\
   \hline
   \T\B  H$_2$  &    0.8829  & $-$3.89 & $-$4.45  & $-$4.45&  $0$ \\
   \T\B  N$_2$  & $-$0.2954  & $-$18.10 & $-$8.95  & $-$8.95& $-$0.23\\
   \T\B  F$_2$  & $-$0.1264  & $-$17.47 & $-$6.83  & $-$6.83& $-$0.13\\
%   \T\B  P$_2$  & $-$0.1165  & $-$42.2137 \red{($-$46.1153)} & $-$14.2575 \red{($-$18.1591)} & $-$14.2575 \red{($-$18.1591)} \\
   \T\B  HF &    0.7419  & $-$10.61  & $-$17.21 & $-$17.21& $-$0.12\\
%   \T\B  CH$_4$ &    0.3526  & $-$19.8081 \red{($-$19.8417)} & $-$19.8081 \red{($-$19.8417)} & $-$19.8081 \red{($-$19.8417)} \\
%   \T\B  C$_6$H$_6$  &          &            &         & \\
     \hline \hline
\end{tabular}
\end{center}
\end{table}

\subsection{Supporting data}

We provide additional data here to complement the numerical results presented in the main text. In Table~\ref{eq20a_sto} we present a numerical test of the sum rule Eq.(15a).
In Table~\ref{eq20bc} we present a validation of the translational sum rules for isolated systems.
Finally, in Table~\ref{tab_magnetic} we report our calculated rotational $g$ factors and magnetic susceptibilities for an additional set of diatomic molecules.

{The $g$ factors present, in some cases, significant deviations with respect to the values that were reported recently in Ref.~\cite{Zabalo-22}.
There are two main differences between the present approach and that of Ref.~\cite{Zabalo-22}: first, the velocity dependence
of the pseudopotentials in the rotating molecule is correctly accounted for in our work; second, our description of the induced magnetic moment is
consistent with the GIPAW method. (In Ref.~\cite{Zabalo-22} the definition of the current density proposed by Ismail-Beigi, Chang and Louie~\cite{Beigi-01}
was used instead.)
The present values are generally closer to the experimental ones, in many cases outperforming the results of (much more demanding) 
explicitly correlated approaches.}

\bibliography{merged}